# Deep learning approach to coherent noise reduction in optical diffraction tomography

## Gunho Choi,[3,6] DongHun Ryu,[1,2,6] YoungJu Jo,[1,2,3,5] YoungSeo Kim,[2,3,4] Weisun Park,[1,2,3] Hyun-seok Min,[3] and YongKeun Park[1,2,3,*]

[1]Department of Physics, Korea Advanced Institute of Science and Technology (KAIST), Daejeon 34141, Republic of Korea
[2]KAIST Institute for Health Science and Technology, KAIST, Daejeon 34141, Republic of Korea
[3]Tomocube Inc., Daejeon 34051, Republic of Korea
[4]Department of Chemical and Biomolecular Engineering, KAIST, Daejeon 34141, Republic of Korea
[5]Present address: Department of Applied Physics, Stanford University, Stanford, CA 94305, USA
[6]These authors contributed equally to this paper
*Corresponding author: yk.park@kaist.ac.kr

We present a deep neural network to reduce coherent noise in three-dimensional quantitative phase imaging. Inspired by the cycle generative adversarial network, the denoising network was trained to learn a transform between two image domains: clean and noisy refractive index tomograms. The unique feature of this network, distinct from previous machine learning approaches employed in the optical imaging problem, is that it uses *unpaired* images. The learned network quantitatively demonstrated its performance and generalization capability through denoising experiments of various samples. We concluded by applying our technique to reduce the temporally changing noise emerging from focal drift in time-lapse imaging of biological cells. This reduction cannot be performed using other optical methods for denoising.

## 1. INTRODUCTION

Recent advances in quantitative phase imaging (QPI) offer an extended opportunity for the label-free, non-destructive, and quantitative study of biological specimens [1]. As a scheme for three-dimensional (3D) QPI, optical diffraction tomography (ODT) is an imaging method that uses angularly varying illumination to reconstruct the 3D refractive index (RI) distribution of a microscopic sample. Since RI, an intrinsic optical property of a sample, provides morphological and biochemical information, ODT has been successfully applied to various fields, including histopathology [2-4], hematology [5-8] microbiology [9], cell biology [10-12] and nanotechnology [13].

The image quality of a reconstructed tomogram can be degraded by the noise originated from the use of coherent illumination (Fig. 1). Unwanted interference of the coherent light generates this noise in the form of fringe patterns and speckle grains [14]. This is mainly caused by multiple reflection from optical elements and dust particles. Misalignment of the optical system could also deteriorate the reconstructed tomogram. We term this category of noise as "coherent noise" throughout this paper.

To remedy the coherent noise, numerous studies involving modifications of experimental setups or additional data capturing have been conducted [15-20]. Unfortunately, this class of methods works only when the imaging system has sufficient stability during measurement. That is, it is challenging to remove the time-varying noises emerging from light source spectrum fluctuations, electro/mechanical vibrations, or focal drifts because of thermal or gravitational effects. Moreover, incoherent ODTs have also been proposed to sidestep the coherence issue, but the practical drawbacks arising from a short coherence length embody dispersion effect and limited angles of illumination [21, 22].

Alternatively, numerical methods that exploit statistical knowledge can mitigate the suppression of the coherent noise, which may address the time-varying noise via post-processing. However, these approaches must assume specific statistics (e.g., Gaussian [23] or Poisson [24]) or need prior knowledge (e.g., sparsity [25, 26]) to enforce the denoising process, which limits the direct application of these techniques to noises of unknown statistics.

In recent years, data-driven approaches, such as deep learning and machine learning, have been a powerful workhorse for various optical imaging problems, including resolution enhancement [27, 28], classification [29-33], in silico fluorescence imaging [34, 35], light scattering [36-38], phase recovery [39], optical system design [40], and noise reduction [41]. With a sufficiently large dataset, a deep neural network embracing non-linear activation functions can approximate any continuous function in the real domain, as first proved by Cybenko's work [42]. Hence, deep neural networks have the potential to design image-to-image transformation models for specific purposes.

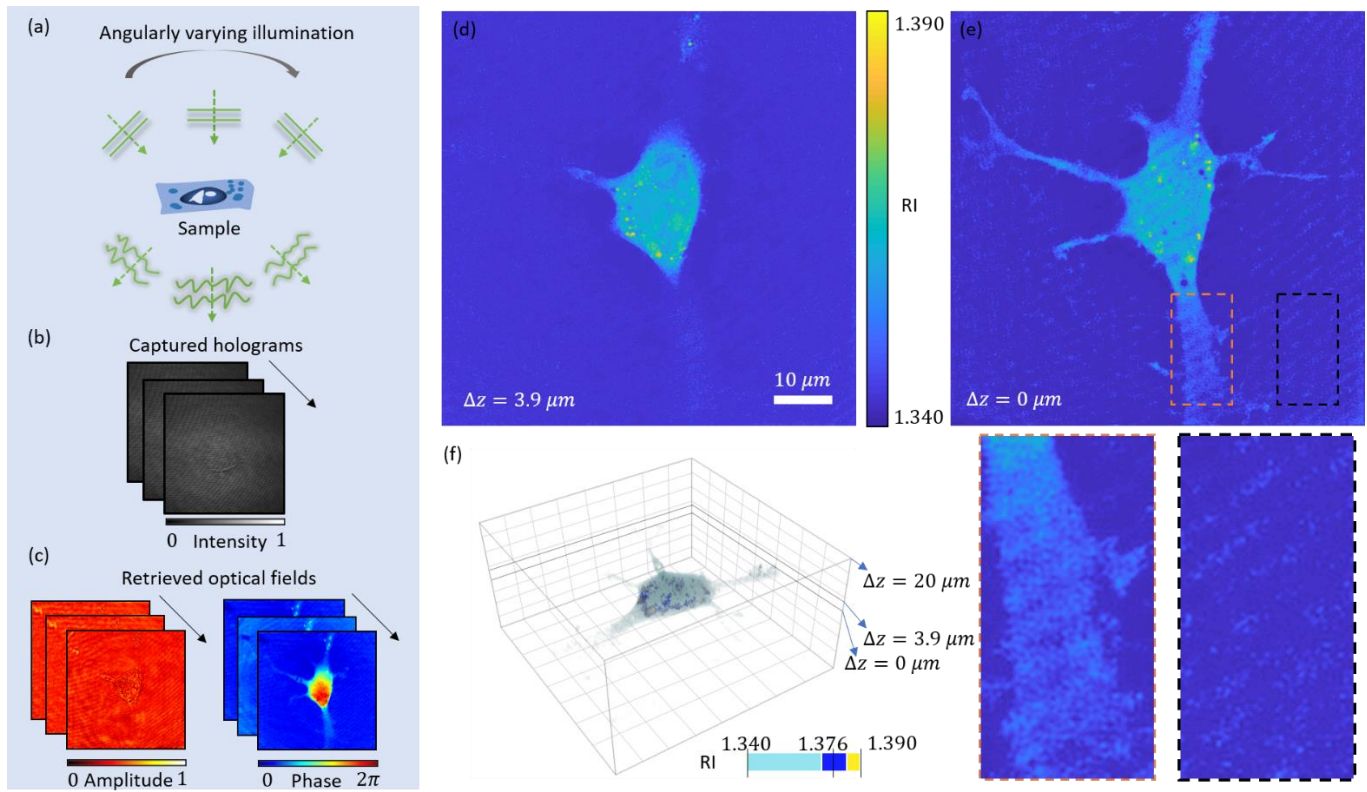

Fig. 1. Coherent noise problem in optical diffraction tomography (ODT). (a-b) The ODT employs angularly varying illumination to capture off-axis holograms. (c) Each complex optical field is reconstructed from the obtained holograms. (d) 2D sliced image of 3D reconstructed tomogram at $\Delta z = 3.9\ \mu m$ e) 2D sliced image of 3D reconstructed tomogram at focus $\Delta z = 0\ \mu m$ corrupted with the coherent noise. (f) 3D rendering of the whole reconstructed tomogram.

A primary disadvantage of the existing networks is the requirement of "image pairs" (e.g., <low-resolution image, high-resolution image>, <brightfield image, fluorescent image>, <brightfield image, phase image>, <speckle pattern, sample image>, and <noisy image, denoised image>). However, in practice, it is often demanding or impossible to obtain such paired training data to use deep learning for denoising tomograms. Obtaining a clean tomogram and the paired coherent noisy tomogram, caused by the thermal focal drift or the inherent system instability, can be difficult and labor-intensive. In addition, preparing such input-output pairs may result in image registration issues.

Our approach for denoising tomograms employs a deep learning framework that takes "unpaired" tomogram sets for training. The deep neural network, inspired by the cycle-generative adversarial network (cycle-GAN), statistically learns to transform between two different image domains (i.e., clean and noisy tomograms) rather than relating one to one in a pair. The trained network was tested to remove coherent noise in the tomograms of silica microbeads for quantitative validation. The performance of the network was also confirmed through several experiments on biological cells never seen by the network during training, demonstrating its generality and potential applicability. Lastly, but most importantly, the denoising network successfully removed the coherent noise in a time-lapse experiment implying a HeLa cell imaging.

## 2. DEEP NEURAL NETWORK FOR DENOISING TOMOGRAMS

The goal of the proposed deep neural network is to learn a high-dimensional function, translating between a noisy image domain $X$ and a clean image domain $Y$, and denoise a 2D laterally sliced tomogram via a trained generator, $G_{XY}: X \rightarrow Y$. Two functions, $G_{XY}: X \rightarrow Y$ and $G_{YX}: Y \rightarrow X$ were trained using two discrimination losses, $L^{D_Y}$, and $L^{D_X}$, where $D_Y$ and $D_X$ are discriminator functions that attempt to discriminate a real image and an image generated by $G$. For instance, $L^{D_Y}$ computes how closely the denoised image $G_{XY}$ (y) follows the true data distribution $y \sim P_{data(y)}$. To enhance optimization convergence, two reconstruction losses, called cycle-consistency losses, were introduced. We minimized the losses comparing an input and a generated image passing through two mirrored functions, $G_{XY}$ and $G_{YX}$. Finally, the trained $G_{XY}$ outputs the denoised tomogram image.

### A. Data acquisition using optical diffraction tomography

We first summarize the ODT reconstruction procedure and the coherent noise for a better understanding of our dataset. Every sample of interest was scanned at various illumination angles to obtain holograms, using a commercialized ODT imaging setup (HT-2H, Tomocube Inc., Republic of Korea), as detailed in the Supplementary Information (SI). Then, 2D optical fields at the sample plane were retrieved from each captured hologram using a field retrieval algorithm exploiting spatial filtering [43]. Following the Fourier diffraction theorem, formulated by Wolf [44], the 3D RI tomogram of the sample was reconstructed. Missing information owing to the limited bandwidth of the system was computationally regularized using a non-negativity constraint [45]. The spatial resolutions in the lateral and axial directions were 110 nm and 360 nm, respectively.

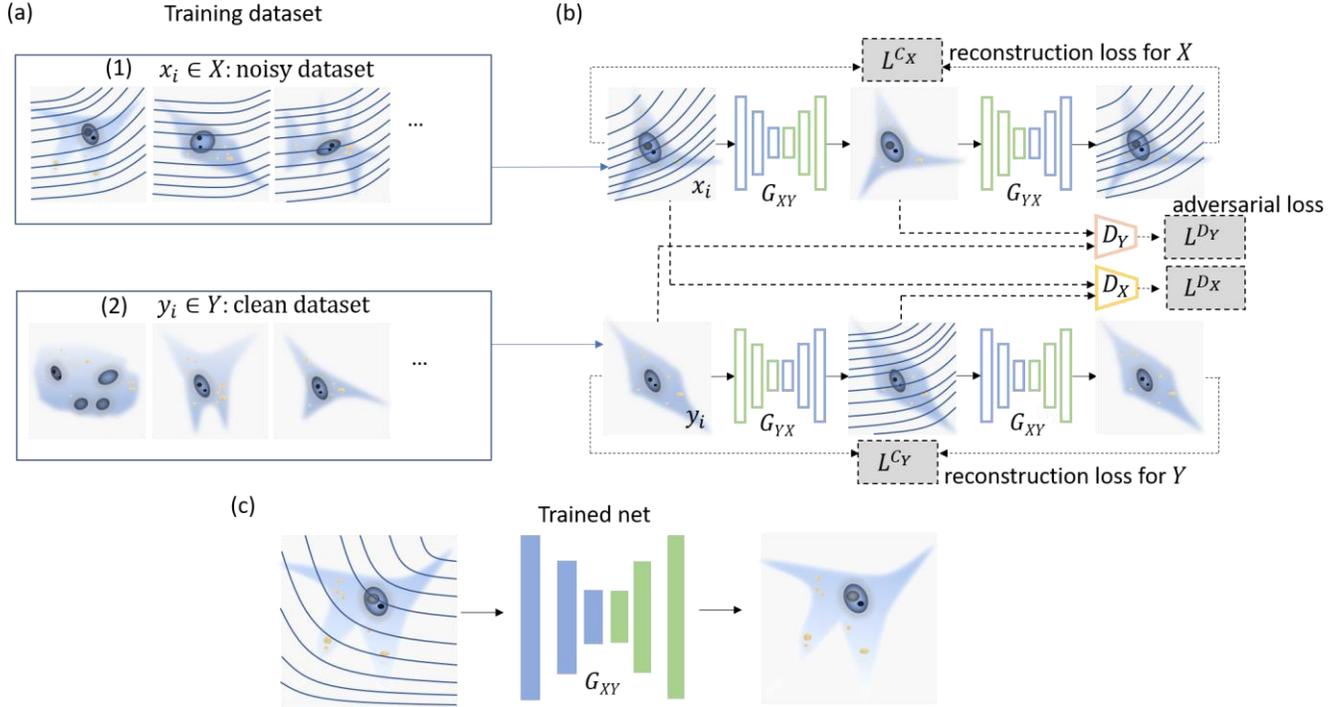

Fig. 2. Overview of the present network for de-noising: training and testing (a) Two classes of dataset for training were prepared. $x_i \in X$: noisy tomogram and $y_i \in Y$: clean tomogram. (b) Training process of the proposed network. $G_{XY}$: Generator that maps $x$ to $y$. $G_{YX}$: Generator that maps $y$ to $x$. $D_Y$: Discriminator to determine if given input is a generated clean image from $G_{XY}$ or a real data $y$. $D_X$: Discriminator to determine if given input is a generated noisy image from $G_{YX}$ or a real data $x$. $L^{D_Y}$: Adversarial loss for $D_Y$. $L^{D_X}$: Adversarial loss for $D_Y$. $L^{C_X}$: cycle-consistency loss for $x$. $L^{C_Y}$: cycle-consistency loss for $y$. (c) Trained network, $G_{XY}$ removes the coherent noise of 2D sliced tomogram.

Two examples of 2D tomograms at different axial depths are shown in Figs. 1(d)–(e). The coherent noise disturbs both the cell features and background in the tomogram at $\Delta z = 0\,\mu m$, while the tomogram slice at $\Delta z = 3.9\,\mu m$ allows to clearly visualize the subcellular organelles. A 3D isosurface image of the reconstructed NIH3T3 tomogram is displayed in Fig. 1(f).

To train the proposed network, we prepared a dataset containing 2D tomograms of NIH3T3 cells using the ODT protocol explained above. The reconstructed 3D tomograms were center-cropped to a dimension of $256 \times 256 \times 100$ voxels. Because the coherent noises, as well as the cells, were not totally spread out along the axial direction, we first attempted to find the focal plane using one of the widely used methods, Brenner gradient [46], and extracted 25 sliced images centered at the determined focus from each 3D tomogram. Then, the 2D sliced tomograms were annotated by two ODT experts and categorized into two sets: (1) $x_i \in X$: noisy tomograms and (2) $y_i \in Y$: clean tomograms (see Fig. 2(a)). We denoted the data distributions as $x \sim P_{data(x)}$ and $y \sim P_{data(y)}$. Again, we emphasize that, in contrast with conventional deep learning frameworks that benefit from one-to-one paired set, two different datasets ($X$ and $Y$) were prepared to train our denoising network. The datasets, $X$ and $Y$ (denoted as (1) and (2) in Fig. 2(a)), finally contained 455 and 5057 tomograms, respectively; there are less noisy images because they are harder to obtain. The rest of the test data, examined below for our model, was composed by tomograms of silica microbead, HeLa, and MDA231 cells. Details about the preparation of the sample can be found in the SI.

**B. Architecture**

To train the datasets, we employed a deep neural network, motivated by cycle-GAN [47], composed of two generator functions, $G_{XY}$ and $G_{YX}$, and two discriminators, $D_Y$ and $D_X$. First, two generators using the U-net architecture learn a statistical model that maps between domain $X$ and domain $Y$. They perform pixel-wise regression. $G_{XY}$ attempts to make a clean image from every input image annotated as a noisy tomogram, $x_i$, while $G_{YX}$ performs exactly the inverse task. Second, two discriminators using the PatchGAN [48] architecture aim to differentiate a real image from an artificial image generated via $G$. $D_Y$ and $D_X$ discriminate between translated images, $G_{XY}(x)$ and $G_{YX}(y)$, and real images, $y$ and $x$, respectively. This architecture is illustrated in more detail in the SI.

**C. Loss Function**

To train the proposed network, we solved a min-max optimization problem with four loss functions. That is, the network was trained such that $G$ and $D$ competed against each other to minimize or maximize the loss functions. These functions were designed to capture, on the one hand, how well $G$ maps one domain into the other and, on the other hand, how well $D$ discriminates between generated and real images. Again, our goal was to train the network through direct competition between $G$ and $D$ so that $G$ could perform the denoising task with enough accuracy after training.

First of all, adversarial losses were applied to both discriminator functions, $D_Y$ and $D_X$. For a function $G_{XY}: X \rightarrow Y$ and the corresponding discriminator $D_Y$, we formulate the loss function as follows:

$$L^{D_Y}(G_{XY}, D_Y, X, Y) = E_{y \sim P_{data(y)}}[\log(D_Y(y)] + E_{x \sim P_{data(x)}}[\log(1 - D_Y(G_{XY}(x))], \quad (1)$$

where $E[X]$ is the expectation of the random variable $X$. $G_{XY}$ aims to minimize this function while the adversarial function $D$ aims to maximize it:

$$\min_{G_{XY}} \max_{D_Y} L^{D_Y}(G_{XY}, D_Y, X, Y). \quad (2)$$

Similarly, when the adversarial loss and the discriminant are $G_{YX}: Y \rightarrow X$ and $D_X$, respectively, the loss function can be formulated as follows:

$$L^{D_X}(G_{YX}, D_X, X, Y) = E_{x \sim P_{data(x)}}[\log D_X(x)] + E_{y \sim P_{data(y)}}[\log(1 - D_X(G_{YX}(y)))], \quad (3)$$

$$\min_{G_{YX}} \max_{D_X} L^{D_X}(G_{YX}, D_X, X, Y). \quad (4)$$

Next, reconstruction losses (cycle-consistency losses) were also employed to further optimize the network. Both generators, $G_{YX}$ and $G_{XY}$, should perform exactly inverse operations in theory, which means that $G_{YX}(G_{XY}(x))$ should return $x$. Hence, we formulated one cycle-consistency loss as

$$L^{C_x}(G_{XY}, G_{YX}) = E_{x \sim P_{data(x)}}[\| G_{YX}(G_{XY}(x)) - x \|_1]. \quad (5)$$

Reversely, the other cycle-consistency loss was introduced as follows:

$$L^{C_y}(G_{XY}, G_{YX}) = E_{y \sim P_{data(y)}}[\| G_{XY}(G_{YX}(y)) - y \|_1]. \quad (6)$$

Finally, we formulated the full loss function consisting of four terms as follows:

$$L(G_{XY}, G_{YX}, D_X, D_Y) = L^{D_Y}(G_{XY}, D_Y, X, Y) + L^{D_X}(G_{YX}, D_X, X, Y) + \lambda \times (L^{C_x}(G_{XY}, G_{YX}) + L^{C_y}(G_{XY}, G_{YX})), \quad (7)$$

where $\lambda$ is the regularization constant that determines the relative importance of preserving spatial features; we set $\lambda = 10$ in this study. We solved the following min-max problem:

$$G^*_{XY}, G^*_{YX} = \underset{G_{XY}, G_{YX}}{\mathrm{argmin}} \, \underset{D_X, D_Y}{\mathrm{argmax}} \, L(G_{XY}, G_{YX}, D_X, D_Y). \quad (8)$$

using gradient-based optimization to obtain the denoising network.

## 3. RESULTS AND DISCUSSION

Here, we optimized our deep neural network and experimentally verified the performance of the optimized denoising network using the ODT imaging system (HT-1H, Tomocube. Inc, South Korea). We imaged and reconstructed them according to the optical setup and procedure detailed in Fig S1. All the reconstructed tomograms had $256 \times 256 \times 100$ voxels, a lateral resolution of 110 nm, and an axial resolution of 360 nm. Thus, the field of view (FOV) of all the 2D sliced tomogram images is 25.6 μm × 25.6 μm.

### A. Model optimization

To optimize the proposed method, we implemented and compared three networks that, as summarized in Fig. 3, differed in the loss function and up-sampling block of the generator. The sample consisted of NIH3T3 cells, which were also used in the training stage. Figs. 3(a1)–(a3) display the original tomogram and two representative subcellular features for comparison. We also examined the Fourier transform of the tomograms.

First, we used the $l_1$ loss function, along with a naïve U-net architecture for the generator (adopted from [49]), to optimize our neural network. It is known that the $l_1$ loss function performs well even in the presence of outliers. Fig. 3(b) displays the denoised tomogram. However, Figs. 3(b2)–(b3) show that the subcellular structure is not resolved and, as marked by the arrows, the clear round feature is missing. Moreover, the checkerboard artifacts are widely dispersed across the denoised tomogram, and they are displayed as a grid in the image (Fig. 3(b1)) and Fourier spectrum (Fig. 3(b4)). The artifacts arise from the overlapped computation of sliding convolution operations. This fact has been intensively studied in [50].

Next, to address the checkerboard artifacts, we used a resize-convolutional U-net [50] with the $l_1$ loss function, as illustrated in Fig. 3(b). This change also led to a better reconstruction result, displaying a better resolution of the guided features but not as clear as in the original image.

Finally, we attempted to employ a structural similarity index loss (SSIM) [51], instead of the $l_1$ function, along with the same resize-convolutional U-net. The result conserves the detailed features of the cell, noted by the arrows, while the coherent noise in the tomogram is eliminated. In addition, the corresponding Fourier spectrum has reduced artifacts in comparison with other models. We adopted this final model as our denoising network, which was thoroughly used to obtain the results reported below.

### B. Quantitative validation

To quantitatively validate the proposed method, we measured the tomograms of silica microbeads ($5\,\mu m$ diameter, 44054-5ML-F, Sigma-Aldrich Inc., USA). A different imaging setup (HT-1H, Tomocube Inc., Republic of Korea), which was not used to obtain the NIH3T3 training data, was utilized for the acquisition of the microbead tomograms to test the generalizability of the present method. As shown in Fig. 4(a), the captured tomograms had unwanted coherent noise, which led to the deterioration of the image quality. The cropped region at the top left corner displays the noise in the form of fringe patterns and speckle grains. The mean value (MV) and the standard deviation (STD) of the RI in this region are 1.3378 and 0.0019, respectively.

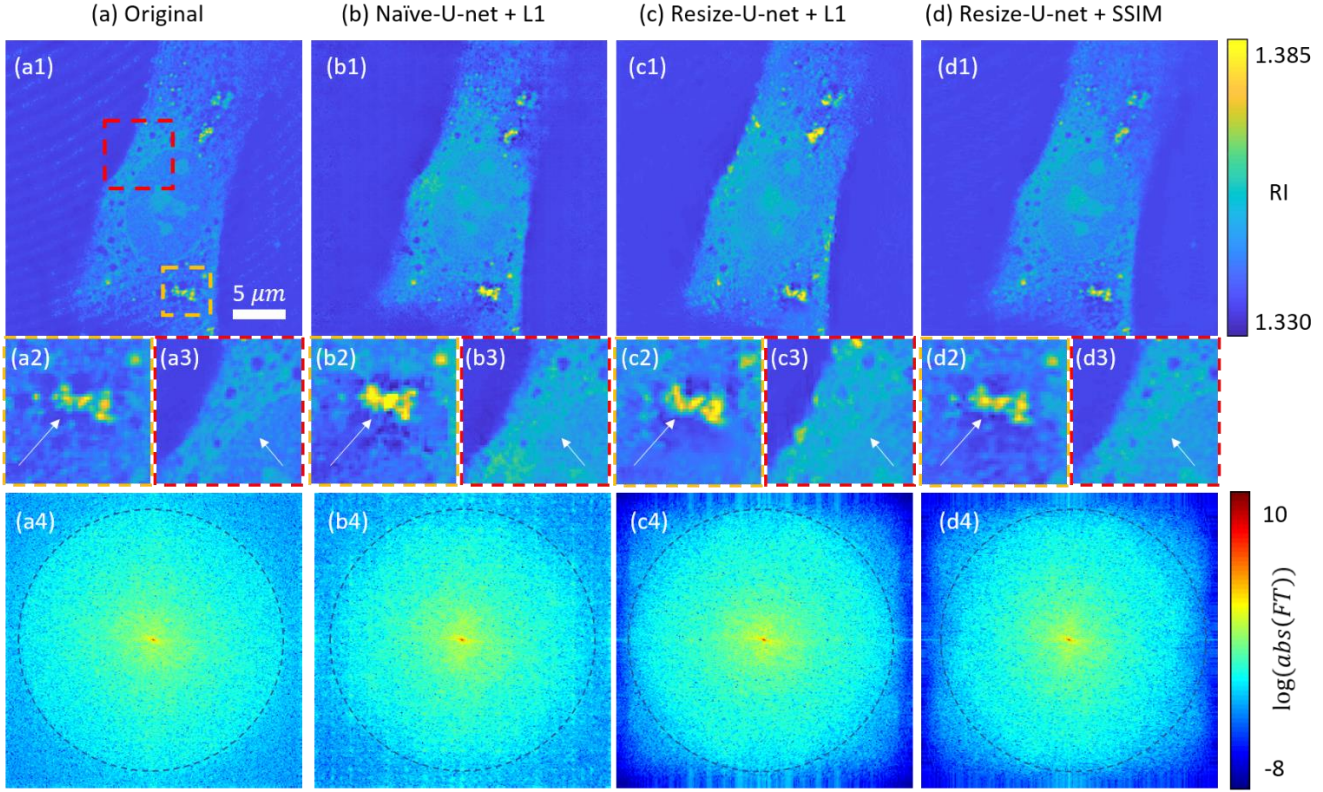

Fig. 3. Architecture search for our deep learning model. (top row) (a) Original tomogram before de-noising. (b) De-noised tomogram image using the Naïve U-net $l_1$ loss function. The detail of subcelluar features is not resolved and the yellow colored artifact is shown, as marked by arrow. The checkerboard effect also appears and can be visualized in Fourier domain. (c) The result of resize-U-net $l_1$ loss improves the image quality by eliminating the checkerboard. (d) To preserve details of spatial features, structure similarity index map loss was utilized, along with the resize-U-net, resolving the subcellular features and clearer boundary of nucleus. (bottom row) Each Fourier spectrum of corresponding tomogram is shown. The black-dotted circle indicates the numerical aperture of imaging system.

Furthermore, as displayed in Fig. 3(b), the proposed method reduces the coherent noise in the tomograms without a significant loss of sample region. Specifically, the generator, $G_{XY}$, trained in the whole network translated the noisy input tomogram into a clear tomogram. Regarding the background region, the MV and the STD of the denoised bead are 1.3369 and 0.0003, respectively, which indicates that, compared to the initial tomogram, the RI values show a better agreement with the theoretical one ($n_{medium}$ = 1.337).

Moreover, the histograms and line profiles for the RI values of the original and denoised tomograms further validate the present method. First, Figure 4(c) displays the data distributions of the background regions enclosed by the blue and orange boxes; a sharper RI distribution can be observed in the denoised tomogram around 1.337. Second, the profiles along the center of the bead demonstrate the consistency of the RI values in the sample region after the application of the present method, as illustrated in Fig. 4(d). Not only is the background region significantly denoised but also the original and denoised RI maps show a close match. It is also noteworthy that the fabrication error of the microbeads makes it challenging to assess the region of the bead accurately.

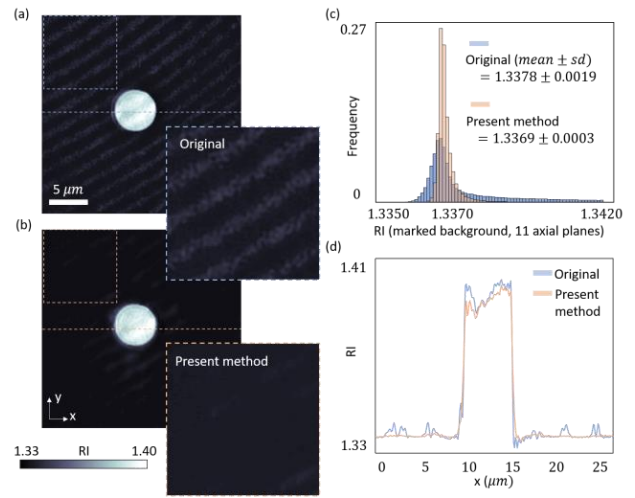

Fig. 4. Quantitative analysis of the proposed network. (a) Original tomogram of the silica microbead degraded by the coherent noise. (b) Tomogram denoised via our method. (c) 2D tomogram slices in the background region (number of slices = 11), marked by top-left corner box, acquired in the axial direction; the RI distributions are shown for comparison to highlight the denoising effect. (d) Line profiles along the horizontal way are visualized.

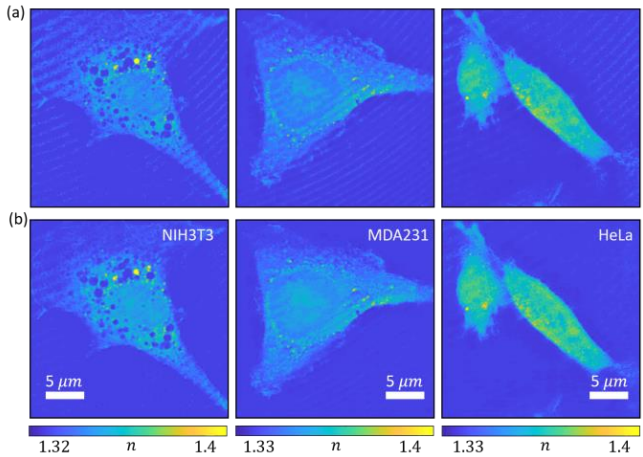

Fig. 5. Experimental validation of the present method. Tomograms of NIH3T3, MDA231, and HeLa (a) in the presence of coherent noise, in the shape of the fringe pattern and (b) after coherent noise removal using our deep neural network.

**C. Demonstration in various biological samples**

To further validate our method, we tested the trained network on various tomograms of eukaryotic cells, including HeLa, NIH3T3, and MDA231. Figure 5 displays the degraded tomograms (first row) and the tomograms denoised using our method (second row). We first applied our network to the pre-split NIH3T3 dataset, which was not used for training. The degraded tomograms, where the coherent noises are clearly visible as fringe patterns, were significantly denoised. As displayed in the second and third column in Fig. 5, we attempted to reduce the coherent noises of the HeLa and MDA231 tomograms to test the generalization capability of our model. As shown in the denoised tomograms, the coherent noises were effectively reduced and the cellular characteristics were conserved in both cases.

Finally, we validated the denoising network through the time-lapse imaging of HeLa cells, as a principal application of our method. We tomographically imaged the HeLa cells for 30 mins with a time interval of 10 mins (see visualization 1). Either a thermal focal drift or a slight change of the specimen could generate a path length difference in the beam path of the ODT imaging system. Hence, as indicated by the orange arrows in Fig. 6 (top row), though we did not modify the imaging system at all, we encountered unwanted noise in the same axial plane, intensified in this time-lapse imaging experiment. In Fig. 6(b), the tomograms denoised using our trained network, which correspond to the images in the first row, are shown. Though the sample information observed slightly changes during the lapse because of the focal drift (e.g., some subcellular or cellular compartments, indicated by black arrows, become faint), the fringe artifacts are effectively eliminated. To more quantitatively compare the original and denoised images, we focused on the noisy and unsampled region of the cell (Fig. 6(c)). Again, we confirmed that the escalating noise level by lapse of time, quantified by the standard deviation, diminished from $0.7 \sim 0.9 \times 10^{-3}$ to around $0.3 \times 10^{-3}$ (Fig. 6 (d)).

## 4. CONCLUSIONS

We have proposed and experimentally validated a deep learning algorithm that suppresses the coherent noise in refractive index tomograms. The deep neural network learns a statistical transformation between two "unpaired" tomogram datasets. We demonstrated its quantitative denoising performance and generalization capability through various biological experiments. Furthermore, in contrast with other optical methods for denoising, the presented method effectively eliminated time-varying noise, generated by thermal focal drift, in time-lapse imaging.

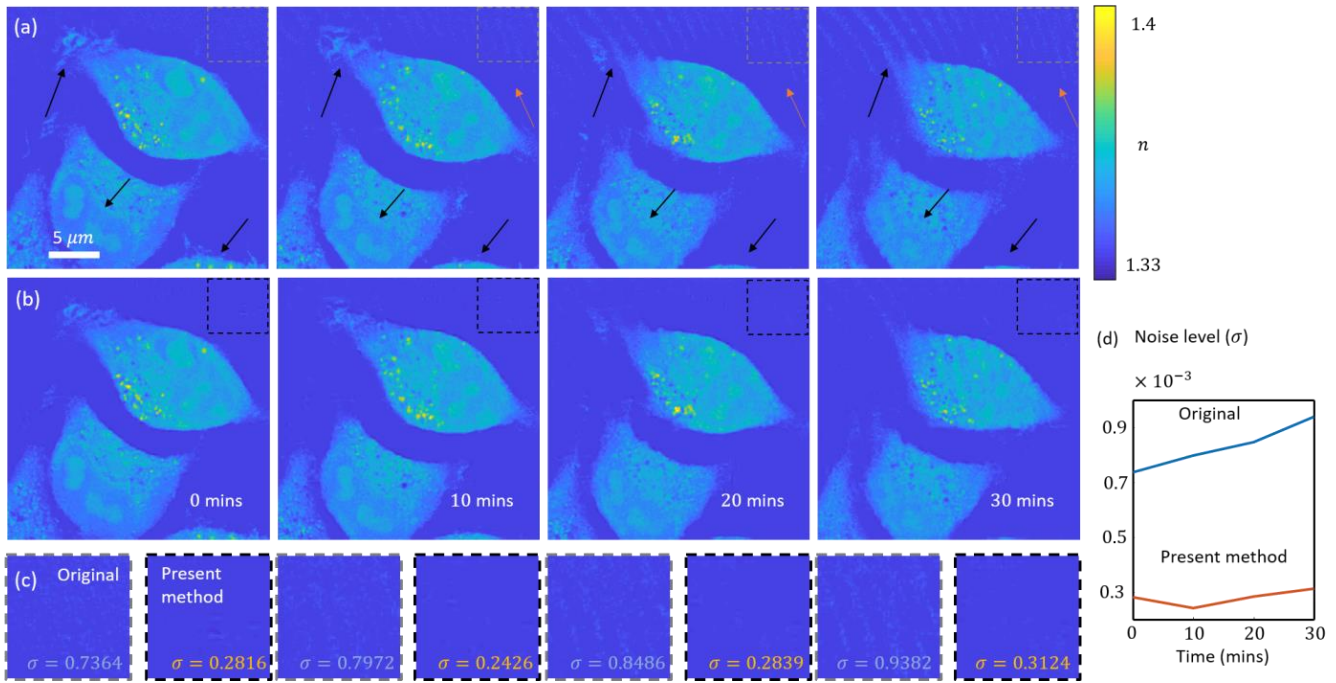

Fig. 6. Time-lapse experiment of HeLa cells with a time interval 10 mins for 30 mins. (a) Original tomograms with incremental coherent noise, induced by focal drift. (b) Our method effectively removes the noises. (c) Comparison of the cropped background regions. (d) Noise level in the regions displayed in (c). It is quantified by standard deviation and decreases upon the application of our method.

One of the primary questions that "data-driven" approaches should answer concerns their generalization ability: can we really suppress the coherent noise in a wide variety of tomograms using a model trained using a specific dataset? We can qualitatively confirm that the denoising performance on NIH3T3 somewhat exceeds the performance on other cells, such as HeLa and MDA231, which have fundamentally different data distribution. We anticipate that the data diversity in the training dataset will improve the generalization ability of the algorithm. This is obviously something to consider to build a better data-oriented model. Secondly, transfer learning may enhance the model for a specific purpose. Instead of retraining the model using every existent dataset, it would be time-efficient to adopt transfer learning (i.e., redesign the architecture of the already trained deep learning model based on an additional dataset).

There are further directions of future work for this paper. First, one can extend the current 2D deep learning framework to a network aimed at denoising whole 3D tomograms using volumetric convolution. The 3D network would exploit additional information from 3D tomograms to help improve the denoising performance (e.g., correlations between different sliced tomograms). Second, though we have here employed a naïve grid-search for parameters optimization, a large number of the hyperparameters and layer designs for training the deep learning algorithm can be tuned using other cutting-edge deep learning technologies, such as reinforcement learning for architecture/parameters search, [cite] to improve the denoising performance. The algorithmic parameters include the number and shape of filters, batch size, optimizer, learning rate, regularization constant, and initialization of convolutional filters. Lastly, we envision that the present approach could be leveraged for the removal of noises belonging to other categories, such as shot noise and Gaussian noise, which prevail in many imaging modalities, including fluorescent imaging, computerized tomography, and X-ray imaging.

**Funding.** This work was supported by Tomocube, and National Research Foundation of Korea (2015R1A32066550, 2017M3C1A3013923, 2018K000396).

**Financial interests.** G. Choi, Y. Jo, Y.S. Kim, W.S. Park, M.H. Min, and Y. Park has financial interest in Tomocube Inc., a company that commercializes ODT and is one of the sponsors of the work.

**Acknowledgement.** The authors thank Soomin Lee and Jiwon Kim for providing the test dataset of the HeLa cells and microbeads. The authors also thank Yoonseok Baek for his constructive review on the manuscript. Y. Jo acknowledges support from KAIST Presidential Fellowship and Asan Foundation Biomedical Science Scholarship.